# ARTICLE

## Self-assembly of small molecules at hydrophobic interfaces using group effect

William Foster[a], Keisuke Miyazawa[b], Takeshi Fukuma[b], Halim Kusumaatmaja[a] and Kislon Voïtchovsky[a]*

Although common in nature, the self-assembly of small molecules at sold-liquid interfaces is difficult to control in artificial systems. The high mobility of dissolved small molecules limits their residence at the interface, typically restricting the self-assembly to systems under confinement or with mobile tethers between the molecules and the surface. Small hydrogen-bonding molecules can overcome these issues by exploiting group-effect stabilization to achieve non-tethered self-assembly at hydrophobic interfaces. Significantly, the weak molecular interactions with the solid makes it possible to influence the interfacial hydrogen bond network, potentially creating a wide variety of supramolecular structures. Here we investigate the nanoscale details of water and alcohols mixtures self-assembling at the interface with graphite through group effect. We explore the interplay between inter-molecular and surface interactions by adding small amounts of foreign molecules able to interfere with the hydrogen bond network and systematically varying the length of the alcohol hydrocarbon chain. The resulting supramolecular structures forming at room temperature are then examined using atomic force microscopy with insights from computer simulations. We show that the group-based self-assembly approach investigated here is general and can be reproduced on other substrates such as molybdenum disulphide and graphene oxide, potentially making it relevant for a wide variety of systems.

## Introduction

Understanding and predicting the self-assembly of molecules into supramolecular structures at interfaces underpins modern material science and is of paramount importance to nanotechnology. Applications range from crystal growth[1] and nanoscale electronics[2] to light harvesting[3] and the nano-functionalisation of interfaces[4] to name only a few examples. Interfacial self-assembly is also ubiquitous in nature, for example in biological processes such the function and folding of biomolecules[5], the formation of cell membranes and protein aggregation[6,7]. Generally, successful self-assembly requires some configurational flexibility for the assembling molecules to be able to probe multiple arrangements, and a stable support or template to assist and stabilise the self-organising molecules. The resulting assemblies are determined by a complex interplay between the interactions present, kinetics, and entropic effects at the interface. This renders any comprehensive understanding of the self-assembly process highly challenging.

The formation of self-assembled structures at solid-liquid or solid-gas interfaces typically occurs in a two-step process whereby molecules first accumulate at the interface and subsequently self-organise into supramolecular structures[8]. During the first stage, the assembling molecules must reside at the interface for relatively long periods of time so as to meaningfully interact with neighbours and promote the relevant ordered structure. At solid-liquid interfaces, this is typically made possible by significant interactions between the assembling molecules and the solid's surface. In systems comprising large molecules, van der Waals interactions can overcome thermal fluctuations[9] and ensure a stable physisorption. However, this becomes more difficult for small molecules (typically < 20 atoms) experiencing lower stabilising forces. Small molecules self-assembly is typically achieved using mobile tethers between the molecules and the solid, but the strategy is inevitably limited to specific molecules and interactions, such as thio-alkanes on gold[10,11]. When specific tethers are excluded, the weak and non-specific surface interactions at play tend to render the self-assembly difficult to understand or predict. Molecular self-assembly in biological systems often rely on such relatively weak interactions in order to create soft or fluid structures that can evolve in response to changes in the environment[12,13]. Yet, biological self-assembly usually occurs at fast rates and with high precision, making it particularly interesting although still poorly understood[14,15].
To date, the self-assembly of small molecules has primarily been studied in extreme cases where systems are under confinement[16] or at low temperatures[17] so as to force the

[a.] Durham University, Physics Department, Durham DH1 3LE
[b.] Nano Life Science Institute (WPI-NanoLSI), Kanazawa University, Kanazawa, Japan.
E-mail : kislon.voitchovsky@durham.ac.uk

molecules to remain long enough near the solid's surface for supramolecular structures to form. Examples in ambient conditions are scarce, with limited insights into the process. This gap is significant because the nanoscale arrangement of small molecules at solid-liquid interfaces is key to phenomena such as friction and lubrication[18], nanomembrane separation[19] and chemical reactivity[20]. Additionally, sophisticated or complex self-assembled structure are likely to involve small molecules as part of their building blocks. The weak dependence of small molecules on specific interactions could also increase the process robustness and flexibility while reducing costs for potential applications.

Recently we have shown that when mixed together, water and methanol, both small molecules, can spontaneously form organised stable supramolecular structures at the surface of graphite (highly orientated pyrolytic graphite, HOPG) at room temperature[21]. This is remarkable because both water and methanol only interact weakly with the hydrophobic HOPG and neither pure solvent can form any stable structure at room temperature. Instead, large heterogeneous self-assembled structures can nucleate thanks to an extended hydrogen bond network that stabilises the assembly by a 'group effect'[20,21]. This result suggests a very different approach to molecular self-assembly: molecules weakly interacting with a solid can be stabilised at the interface by a network of inter-molecular interactions already existing in the liquid[22] albeit transiently. In this framework, the surface mainly serves to reduce the configurational entropy and mobility of the molecules for the self-assembly to begin. This approach is particularly well-suited to small molecules able to form hydrogen bonds.

Here we propose to exploit this platform to explore some of the main factors influencing group-based self-assembly, in particular the interplay between molecule-molecule and molecule-surface interactions. First, the fact that weak solid-liquid interactions are at play should allow added molecules able to interfere with the molecule-molecule interactions to affect the resulting self-assembled structures. In principle, only trace amounts of these added molecules could already have a significant impact if the assembly relies on group-effect. Second, the self-assembly process should not strongly depend on the choice of solid, hence increasing the generality and applicability of this self-assembly mechanism.

Starting from a simple water-methanol system at the interface with HOPG, we first modify the methanol-water interactions by adding small amounts of foreign molecules able to locally modify the hydrogen bond network. Second, we change the ratio of methanol to water, also in conjunction with foreign molecules. Third, the methanol is replaced with primary alcohols exhibiting progressively longer backbones to increase the relative importance of interactions with the substrate in the self-assembly process otherwise dominated by the interfacial hydrogen bond network. Finally, we explore the generality of the self-assembly process by replacing HOPG with molybdenum disulphide ($MoS_2$) and graphene oxide (GrO). $MoS_2$ is a non-organic hydrophobic solid whereas GrO is weakly hydrophilic and has a less regular surface than HOPG. We use primarily high-resolution atomic force microscopy (AFM) to examine the nanoscale details of the different interfacial assemblies. Wherever possible, we complement the experimental findings with Molecular Dynamics (MD) computer simulations.

## Results

### 1. Adjunction of small 'influencer' molecules

We first explore the ability of added foreign molecules - hereafter referred to as 'influencers' for simplicity- to modulate the molecular arrangement of methanol-water structures at the interface with HOPG. In principle, countless molecules can be used as influencers. Here we decided to use some of the constitutive molecules and ions of the standard laboratory buffering agent for biological systems: phosphate buffered saline (PBS). We compare systems containing the pure solvents with that doped with small amounts of PBS, or its main components in isolation, namely disodium phosphate ($Na_2HPO_4$), sodium chloride (NaCl) and potassium chloride (KCl) (Fig. 1). We therefore explore five aqueous solutions: (i) ultrapure water, (ii) a PBS solution comprising 10 mM $Na_2HPO_4$, 137 mM NaCl and 2.7 mM KCl (hereafter simply referred to as PBS solution), (iii) a 10 mM $Na_2HPO_4$ aqueous solution, (iv) a 137 mM NaCl aqueous solution and (v) a 2.7 mM KCl aqueous solution. In all cases, the aqueous solution is mixed 50:50 by volume ratio with methanol and the resulting mixture placed in contact with HOPG. High-resolution amplitude modulation atomic force microscopy (AFM) in liquid [23] is used to explore *in-situ* the sub-nanometre details of the resulting interfacial molecular assemblies. When operated in solution and at small amplitude, the technique can routinely provide molecular-level details of the interface[24,25], including of solvation structures[26,27] and stable supramolecular assemblies[21]. The AFM results include simultaneously acquired topographic and phase images of the interface. Topographic images are often preferred for their obvious interpretation but phase images are sensitive to variations in the interactions experienced by the scanning tip and can hence be used to complement topographic information[28,29].

For each system, we conducted the experiment both with and without the methanol in order to ensure that any molecular assembly observed does indeed involve both types of molecules. Additionally, the AFM data was always collected within an hour of the liquid deposition onto the surface. The is because micromolar quantities of methanol can be produced directly at the HOPG-water interface[20]. The energetics of the underlying mechanism is still unclear, but it typically occurs on longer time-scales, with hours needed to create quantities able to nucleate heterogeneous interfacial structures. By restricting our observations to less than an hour, the influence of methanol produced in-situ can be ignored, as confirmed by the controls in Fig. 1.

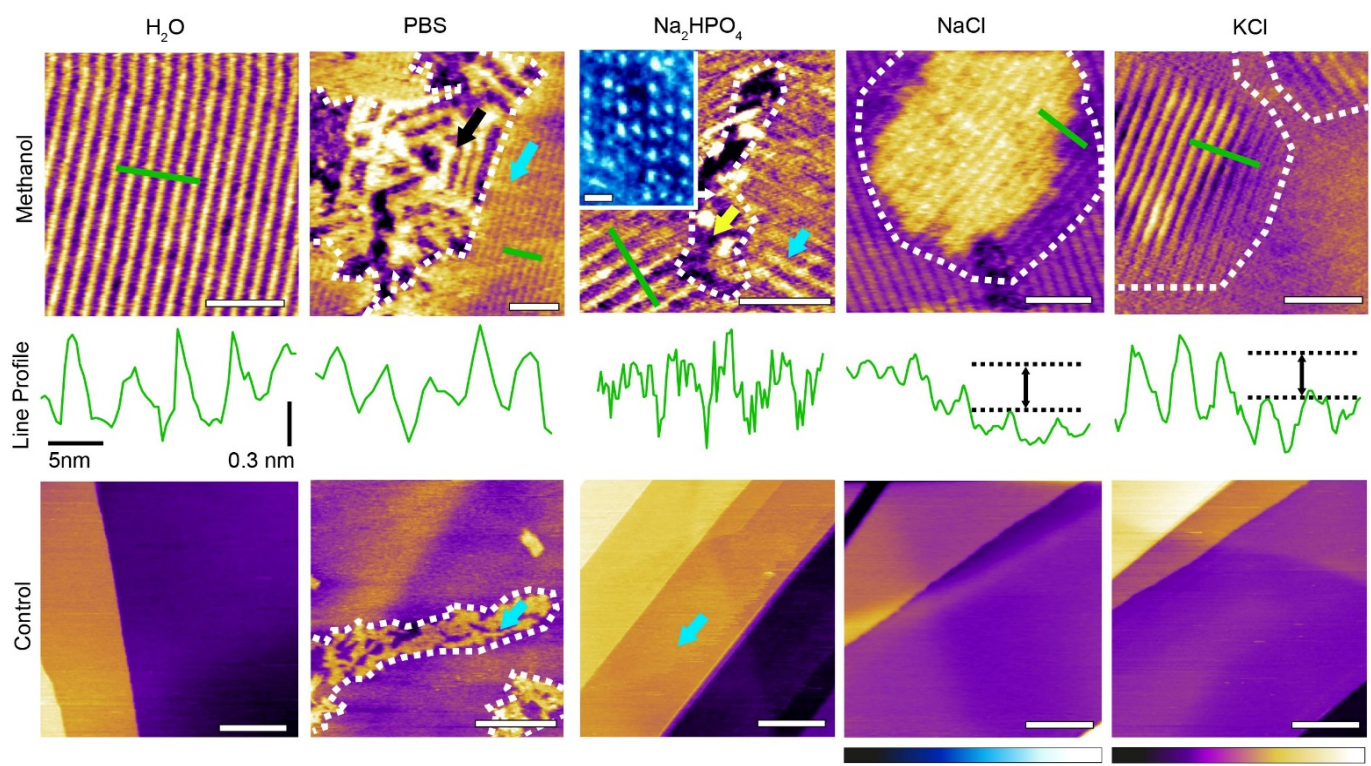

**Figure 1:** Representative AFM images of the HOPG-liquid interface in solutions containing different influencers. The solutions contain a 50:50 volume ratio mixture of methanol and the aqueous component (see text), expect for the controls that contain only the aqueous component. When methanol is present in the solution (top), characteristic methanol-water longitudinal rows[20,21] can be observed with inter-row periodicities of 5.1 ± 0.2 nm, as shown by the corresponding line profiles beneath each figure. The green bars on the images represent the location of the line profile. Swapping the pure water with PBS (10 mM $Na_2HPO_4$, 137 mM NaCl and 2.7 mM KCl) creates complex domains comprising the row-like structures (light blue arrow) and individual ribbon-like structures (black arrows). When only the buffering agent is present (10mM $Na_2HPO_4$) different structures can also co-exist (yellow and light blue arrows) with new features covering only a small area. The inset shows a magnified phase image over these new structures. When only the metal ions are present in the aqueous solution (137 mM NaCl or 2.7 mM KCl), the longitudinal rows visible in the methanol-water mixture re-appear, but some exhibit an altered z-profile with an upward shift of the rows by 0.30 ± 0.06 nm (black arrow). Control experiments conducted in the absence of methanol exhibit no clear assemblies, with only transient features visible in $Na_2HPO_4$ and PBS (blue arrows). The scale bars represent 25 nm (top), 2.5 nm (inset) and 100 nm (controls). All images are topographic images except for the inset (phase). The colour scale bars represent 0.5 nm height variation for the methanol-water mixture on the top row, 1nm for the rest of the top row, and 3nm for the bottom row images except for the PBS (1nm). The blue scale bar represents a phase variation 10°. The data shown was obtained with the AFM tip fully immersed in the solutions, and within 1 hour of depositing the liquid droplet onto the HOPG surface.

When water and methanol are both present in the solution, molecularly ordered monolayers nucleate. They typically consist of ordered row-like features with an inter-row periodicity of 5.30 ± 0.20 nm, consistent with previous reports[20,21]. High-resolution images of these rows reveals ångström-scale perpendicular sub-features (see Supplementary Fig. 1) corresponding to water and methanol molecules forming correlated parallel 'wires' on the surface of HOPG. This arrangement has been shown to offer one of the lowest energy configurations for the mixed molecules[21]. It will hereafter be referred to as the 'basic methanol-water monolayer'.

Using the PBS solution as an influencer induces the co-existence of two different structural domains: the basic methanol-water monolayer (Fig. 1, blue arrow) and regions presenting a new type of ribbon-like structures (black arrow). Over the course of a typical experiment, both structures occupy a comparable area, and remain unperturbed by the scanning AFM tip. The ribbon-like features exhibit similarities with the basic structure suggesting the presence of methanol in the assembly. However, the irregular periodicity (black arrow) point to a significant impact of the influencers on the assembling methanol and water structures. Understanding the precise role of the influencers is however challenging due to the PBS solution containing three types of molecules at different concentrations. To better assess the role of these component, we investigated separately each of the PBS components at their PBS concentration in water-methanol mixtures. When only the buffering agent, $Na_2HPO_4$, is present in the aqueous solution, fine rectangular patterns appear at the boundary between basic methanol-water monolayer domains (Fig. 1 inset). The area covered by these features is comparatively small and the pattern is easily deformed or damaged by the scanning AFM tip (Supplementary Fig. 2). This indicates weakly bound structures compared to the basic water-methanol motif. When dissolved in water, $Na_2HPO_4$ disassociates into sodium ($Na^+$) and phosphate ($HPO_4^{2-}$ and $H_2PO_4^-$) ions, the latter being able to form multiple hydrogen bonds. Here the fine structures observed suggest phosphate ions to have been incorporated within the methanol-water assembly. Molecular Dynamics (MD) simulations cannot provide detail insights into the molecular arrangement of the interface given the weak interactions at play[21], but bulk MD simulations suggest that the phosphate and sodium ions form clusters with their hydrogen bonding groups facing outwards towards the surrounding liquid (Supplementary Fig. 2), and should therefore able to be incorporated into the basic structures. A similar behaviour has been previously reported for calcium diphosphate[30]. At very low concentrations these clusters are comparable in size to the dotted features observed (inset) which may be explained by hydrated ionic clumps incorporated into the basic monolayer. This would also be consistent with recent reports of long residence times for metal ions at hydrophilic interfaces, here the monolayer[24,31].

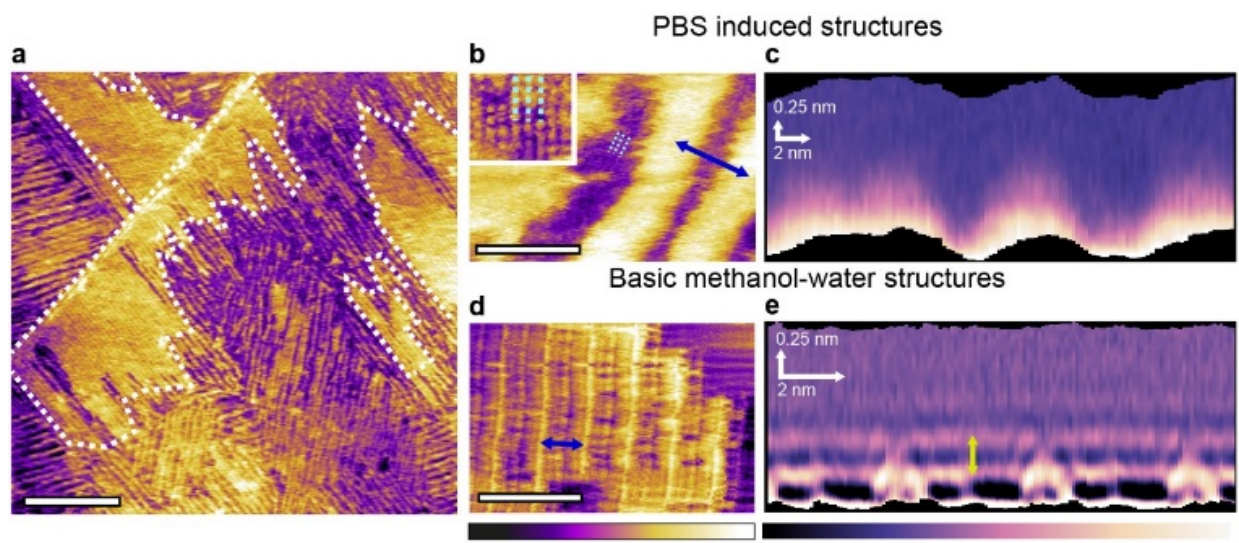

**Figure 2:** Influence of PBS on the 3-dimensional self-assembly of water-methanol structures at the interface with HOPG. Frequency modulation imaging of the interface between HOPG and a 50:50 mixture of methanol and PBS solution shows the two types of domains already visible in Fig. 1. The boundary of the taller domains is highlighted by the white dashed line (a). At higher magnification, domains unique to the methanol PBS mixture exhibit features running parallel to the rows (inset and dashed lines) (b). These fine features exhibit a periodicity of 0.94 ± 0.06 nm. Taking a 3D SFM cross-section horizontally across the taller rows in (b) does not reveal any particular solvation features when moving vertically away from the interface (c). For comparison, the same analysis conducted in a 50:50 mixture of methanol and ultrapure water yields the basic monolayer, here with a row spacing of 4.65 ± 0.08 nm (d). A 3D SFM cross-section analysis (e) reveals clear hydration layers with a vertical periodicity of 0.30 ± 0.05 nm (yellow arrow). The scale bars are 100 nm in (a) and 10 nm in (b) and (d). The purple-yellow colour scale bar represents a topographic variation of 0.5 nm in (a), 0.6 nm in (b) and 0.3 nm in the inset and 0.4 nm in (d). The purple-white scale bar represents a frequency shift variation of 2 kHz in (c) and 3 kHz in (e).

Using only the metal salts (NaCl or KCl) does not impact the lateral order of the basic methanol-water monolayer. Instead the metal ions appear to induce an upward shift of the rows by 0.30 ± 0.06 nm from the average height of the basic methanol-water structure (black arrow on line profile, Fig. 1). A similar vertical displacement could also be observed for the system containing the Na$^+$ ions from Na$_2$HPO$_4$ (see Supplementary Fig. 3). Vertical stacking of multiple basic methanol-water monolayers has previously been observed in pure water-methanol mixtures[21], but this is unlikely to be the case here as evident from the continuity of the rows in the profiles. Enhanced resolution images on the raised structures in the NaCl experiment show features that we interpreted as molecular clusters involving the metal ions and residing on top of a basic row structure (see Supplementary Fig. 4). These clusters tend to follow the pattern of the underling rows. This view is compatible with previous experiments where electric fields were used to reversibly adsorb metal cations or anions on top of the basic structures[21], thereby inducing raised row-like structure in registry with the underlying methanol-water assembly. The metal ions themselves cannot form hydrogen bonds and sitting atop the assembly would allow them to remain fully hydrated while altering the local hydrogen bonding properties. The raised patches therefore likely result from this perturbation to the local hydrogen bond network. The non-raised regions in Fig. 1 appears smooth and regular, with no evidence of clusters.

Control experiments show that ultrapure water itself is unable to form stable structures on HOPG at room temperature[21] (Fig 1). This is also the case for pure methanol[32]. The absence of any interfacial structure in ultrapure water and in the salt solutions is fully expected. Larger ions such as Na$_2$HPO$_4$ and to a larger extent in the PBS solution are able to form faint disordered layers at the surface of HOPG, but no molecular ordering is visible.

Interestingly, changes to basic water-methanol structure in the presence of multiple influencer molecules (PBS) appear significantly more pronounced than could be expected from a simple addition of changes observed in the individual components at equivalent concentrations (Na$_2$HPO$_4$, NaCl and KCl). This points towards a complex interplay between the influencers and the hydrogen bonded networks stabilising the system. Experiments conducted with the individual molecules suggest the incorporation of the phosphate ions into the basic monolayer assembly, and the ability of the metal ions to shift the monolayers despite their lack of direct hydrogen bond. One plausible explanation for this cooperative behaviour of the influencer is that once the phosphate ions become involved in the hydrogen bonded networks of the monolayers, their charged nature encourages interaction with metal ions, allowing the latter to have a larger influence on the resulting structures.

To gain further insights into the hydration properties of the stable new structures observed in PBS, we repeated the experiments using small-amplitude frequency-modulation AFM. This operating mode, while potentially more challenging on soft samples, enables precise 3-dimensional mapping of the liquid density near the interface[33–35]. This makes it possible to derive in-situ local quantification of the structures' thickness and shape in three dimensions, when moving away from the HOPG surface (Fig. 2). The technique is often referred to as 3 dimensional scanning force microscopy or 3D-SFM[36]. The size of the domains formed in PBS and their stability under imaging conditions makes the system suitable for investigations with 3D-SFM.

Frequency modulation AFM is able to resolve both the basic methanol-water assemblies and the PBS-specific longitudinal domains in solution (Fig. 2a). The former structures are characteristically highly ordered and periodic, whereas the latter exhibit a significant degree of variability in the periodicity of the features (see also Supplementary Fig. 5). Higher resolution images of the PBS-induced structures (Fig. 2b) reveal molecular-level features running parallel to the main rows. A representative 3-dimensional section taken over the PBS-specific interfacial domains (Fig. 2c) shows no clear solvation features other than the rows themselves. In contrast, when the same analysis is performed on the basic methanol-water rows formed in the pure solvents (Fig. 2d-e), intricate solvation features extend in the bulk solution, with multiple well-defined hydration layers (spacing of 0.30 ± 0.05 nm) visible directly above the basic monolayer. The inter-layer spacing is similar to that reported for the HOPG-degassed water interface[37] and simulations of a HOPG-water-methanol interface[21], suggesting little direct interaction between the basic methanol-water monolayer and the interfacial liquid. This is consistent with the molecular model of the basic monolayer where all the available hydrogen bonds are engaged within the layer[21], leaving little to interact with the surrounding solvent. Interestingly, the layer spacing is smaller than the ~0.5 nm spacing reported for hydration layers at the non-degassed water-HOPG interface[37,38]. This larger spacing was attributed to dissolved molecules displacing the water from the interface with HOPG. The present observations suggests that the structured basic monolayers is able to prevent such a displacement of water to occur.

## 2. Changes in the alcohol-water ratio

The data presented in Figs. 1 and 2 demonstrate that influencer molecules can alter the self-assembly, leading to a variety of different structures that can differ substantially from the basic methanol-water monolayer, both in morphology and in their local interaction with the surrounding interfacial liquid. In all cases, water and methanol are both needed for well-ordered structures to nucleate. However, their respective molecular proportions can be changed, offering an additional route to influence the interfacial self-assembly, especially when in the presence of influencers. To illustrate this point, we varied the methanol-PBS ratio from Fig. 2, reducing the methanol concentration down to 5%. This results in the formation of intricate, highly ordered structures with a rectangular lattice (Fig. 3a). These structures are reminiscent of those visible in a 5:95 mixture of methanol and ultrapure water (Fig. 3b), but the respective unit cell in each system differs in shape and size, once again highlighting the specific role played by PBS in the interfacial molecular organisation.

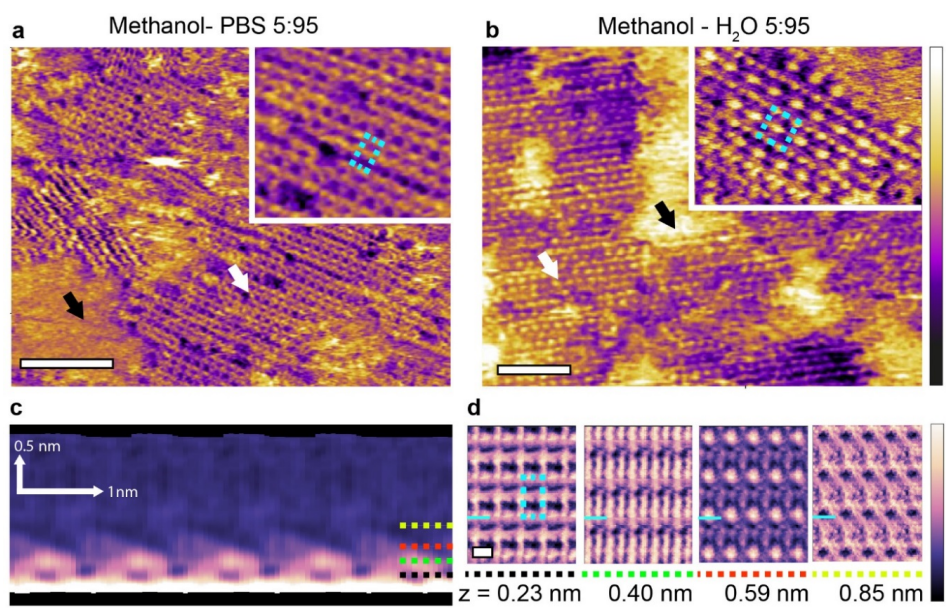

**Figure 3:** AFM imaging of the unique structures produced in a 5:95 mixture of methanol with the PBS solution on HOPG. Two types of domains are visible (a) (black and white arrows). The inset shows a magnified view of the fine structured region (white arrow), indicating a unit cell of 0.90 ± 0.05 nm by 0.82 ± 0.05 nm (blue dashed line). For comparison, images acquired in a 5:95 mixture of methanol and ultrapure water (b) also show some fine structure across ordered (white arrow) and disordered (black arrow) regions, but with a different unit cell (inset, blue dashed line). A 3D SFM cross-section taken perpendicularly to the features denoted by the white arrow in (a) reveals complex 3D motifs that extend up to 0.85 nm in the vertical (z) direction (yellow dashed line) (c). These motifs are best visualised by taking horizontal cross sections parallel to the HOPG surface at different heights (d). In all cross sections, the blue line indicates the position where the vertical cross section shown in (c) was taken. The rectangular unit cell from (a) is overlaid on the lowest of the four horizontal cross sections. Images in (c) and (d) have been processed with an average filtering process that uses a pattern matching algorithm. Details of the procedure are described in a previous work[34]. The raw data is given in Supplementary Fig. 6. The scale bars are 10 nm in (a) and (b) and 1 nm in (d). The purple-yellow colour scale bar represents a height variation of 0.2 nm in (a) and the inset, 0.8 nm in (b) and 0.5 nm in the inset. The purple-white scale bar represents a frequency shift variation of 0.8 kHz in (c) and 0.1 kHz in (d).

Three-dimensional SFM mapping of the interface between HOPG and the methanol-PBS system reveals periodic features extending up to 0.85 nm away from the surface of HOPG (Fig. 3c). The associated solvation structure is remarkably intricate with a lateral pattern changing dramatically at different distances from the surface (Fig. 3d). The transition from the 2D monolayers observed in the 50:50 mixtures of methanol and PBS to the 3D structures observed in the 5:95 mixture may be in part explained by comparisons with the hydrogen bonding behaviour of methanol-water mixtures in the bulk: at low methanol concentration, numerous experimental and computation studies[22,39–41] have demonstrated that three dimensional hydrogen bonded structures dominate due to the tetrahedral coordination of water. In contrast, one- and two-dimensional hydrogen bonded structures such as chains and rings dominate at higher methanol concentrations. Here this could explain why three-dimensional solvation structures develop from the interface at low methanol concentration (Fig. 3 c, d) whereas linear features in the basic methanol-water monolayer are predominant at higher alcohol concentrations (Fig. 1 and 2). The exact effect of the PBS is harder to pinpoint. No visible hydration layers were observed above the interfacial structures developing when PBS is present (Fig. 2), suggesting a higher degree of similarity with the two-dimensional assemblies. It should be pointed out that possible tip effects on the 3D-SFM measurements cannot be ruled out, but such effects would be similar on all 3D results. Yet, there still remains clear solvation differences between the 5:95 and 50:50 methanol-aqueous solution mixtures as well as in the presence of PBS (Figs. 2c, e, and 3c).

## 3. Tuning of surface interactions through the length of the alcohol backbone

These last results confirm that influencers and the ratio of alcohol to water can both affect the interactions between the different liquid molecules at the interface and hence the resulting supramolecular structures. There exists a third, more fundamental route to influence the self-assembly process: the strength of the interaction between the assembling molecules and the substrate. To enable self-assembly by group effect, this interaction must remain relatively weak compared to thermal fluctuations. Stronger interactions will tend to bring the system back to the traditional 2-step self-assembly regime. If the interaction strength can be tuned, the relative importance of inter-molecular forces and substrate effects can be controlled. In water-alcohol mixtures, this is tuned by the length of the alcohol's alkyl backbone: the longer the carbon backbone, the stronger the attraction to HOPG.

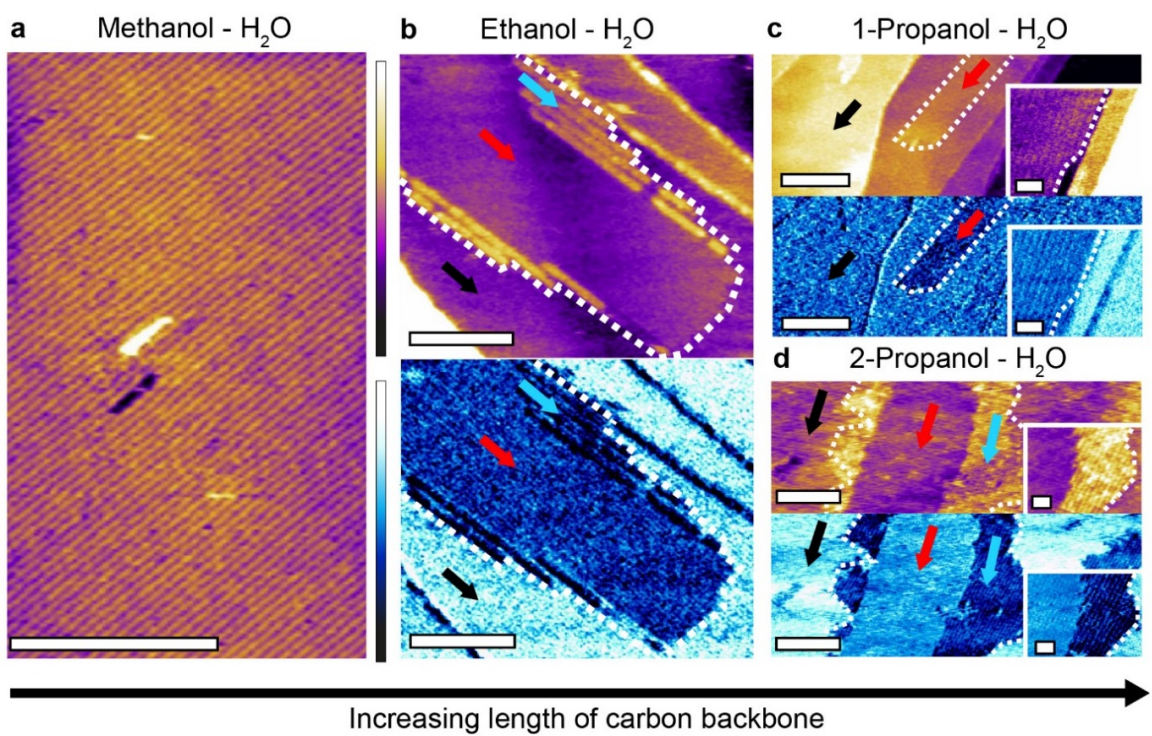

**Figure 4**: Impact of the backbone length of primary alcohols on interfacial self-assembly on HOPG. The basic monolayer motif is visible as expected in a 50:50 methanol:water mixture (a), here imaged by amplitude-modulation AFM (topography image). In a 50:50 ethanol:water mixture (b), two organised layers are visible both in topography and in the phase where it is more pronounced, outlined by a white dashed line (blue and red arrows). In phase, the self-assembled layers appear darker than the directly exposed graphite, where no structures are present (black arrow). The lower layer shows few resolvable features and is bordered by wide rows that have a separation of 5.89 ± 0.28 nm. In 50:50 1-propanol:water mixture (c), novel structures with long, straight edges emerge (red arrow) and grow on top of the exposed graphite (black arrow). The structures have a row periodicity of 5.86 ± 0.25 nm. The inset shows details of the longitudinal row structures near an edge. Further variance is seen in a 50:50 2-propanol:water mixture (d) where two types of domains form (red and blue arrows), both demonstrating a clear phase contrast with the graphite surface (black arrow). The domains have longitudinal rows with periodicities of 6.10 ± 0.35 nm (blue arrow) and 4.91 ± 0.45 nm (red arrow). Unlike for (c), higher resolution of the row (inset) evidence curved edges. The scale bars are 50 nm in (a) and (b), 100 nm in (c) and (d) main image and 20 nm in the insets. The purple colour scale bar represents a height variation of 1 nm in (a), (b) and (d), 3nm in (c) and 0.5 nm in the insets. The blue scale bar represents a phase variation of 1.5º in (b), 2º in (c) and its inset and 15º in (d) and its inset.

To systematically investigate this effect, we compared the interfacial structures formed in binary mixtures of ultrapure water with alcohols presenting increasingly longer carbon chains such as ethanol and propanol. We initially chose primary alcohols due to their topological similarity to methanol, which allows for a direct comparison with the water-methanol monolayers. Selected results comparing the characteristic water-methanol assembly molecular structures observed in mixtures of containing ethanol ($C_2H_6O$) and 1-propanol ($C_3H_8O$) are shown in Fig. 4.

It is immediately clear that more complicated linear structures can form in the presence of longer alcohols. In a 50:50 mixture of ethanol and water two different types of molecular arrangements are visible (Fig. 4b). A uniform layer (red arrow) with a height of 0.24 ± 0.05 nm above the HOPG surface is partially covered by a second layer 0.62 ± 0.05 nm high and composed of row-like structures. A clear phase difference is visible between the structures and the HOPG, confirming distinct molecular arrangements. Repeat experiments in the ethanol-water mixture revealed other types of structure, often with periodic row-like features exhibiting sharp domain edges that are uncommon in methanol-water mixtures (Supplementary Fig. 7).

Increasing the carbon backbone length further and using 1-propanol-water mixtures induces a novel type of structural domain (Fig. 4c) which also exhibits straight edges, similar to those formed in ethanol-water mixtures (see Supplementary Fig. 8 for a comparison). These highly linear domains are unstable under imaging conditions and disassemble within minutes (see Supplementary Fig. 9) indicating the size of the assembling molecules is starting to hinder their ability to form extended hydrogen bonded networks.

Generally, the more elongated, sharp-edged domains observed with larger primary alcohols suggest a stronger epitaxial effect, consistent with the marginally increased alcohol-graphite interaction. This is most obvious for the 1-propanol-water mixture. The fact that only linear structures are observed indicates a predominant role of the one-dimensional molecular chains associated with primary alcohols[22,42–44]. Indeed, structures become less linear when 1-propanol is replaced with 2-propanol (Fig. 4d), with two novel competing ordered domains appear, exhibiting more frayed and rounded boundaries. The results in 2-propanol also highlights the flexibility of the interfacial hydrogen bonded network, including their ability to incorporate molecules with differing shapes and structures.

The largest primary alcohol still able to mix with water is 1-hexanol ($C_6H_{14}O$) with a solubility limit of around 0.7% in ultrapure water. With a carbon backbone twice as long as 1-propanol, the interaction between 1-hexanol and the HOPG surface is significantly stronger in water, thereby offering an ideal system to test the limit of hydrogen bond-based group stabilisation as opposed to traditional surface bound self-assembly. Pure n-hexanol naturally forms ordered structure at the surface of HOPG (from vapour) at temperatures below -10 °C [9] whereas shorter alcohols require significantly lower temperatures to form ordered structures in similar experiments (e.g. -130 °C for methanol[45]). When at its solubility limit in water, 1-hexanol forms self-assembled structure with several different features that can be resolved with molecular resolution (Fig. 5a). Certain areas of the sample retain structures comparable in shape and size to the basic methanol-water monolayer (Fig. 5b). Given the low concentration of hexanol, the small amounts of methanol naturally produced by water catalysis[20] may be responsible for creating basic monolayers. However, the presence of hexanol appears to destabilise these structures which can easily be removed from the surface with the AFM tip to expose the HOPG underneath (Fig. 5c). This is an unusual behaviour for the basic monolayers and suggests that the system is being disrupted by the addition of 1-hexanol to the point where the intermolecular hydrogen bonds are no longer sufficient to stabilise the monolayers. The system appears to have finally reach the point where direct molecule-substrate interactions can seriously compete with hydrogen bonded molecular networks to drive and control the self-assembly. This conclusion is also backed by MD simulations of the water-hexanol system.

We ran MD-simulations of a system consisting of 10:45:45 hexanol-methanol-water mixture at the interface with 8 layers of graphite in a super-cell geometry. The total number of atoms is approximately 22,000 and the simulation covered 16 ns (Fig. 5d, see methods for more details). The relatively high concentration of hexanol was chosen to reflect its expected increased concentration at the interface with HOPG when compared to bulk concentrations[21,46,47]. The presence of methanol accounts for the fact that small quantities of methanol are always present at the interface due to in-situ catalysis of water[20], and may play a role here due to the relatively low hexanol bulk concentration at saturation (0.7%). The simulations show the formation of a self-assembled solid-like layer of molecules dominated by the hexanol (Fig. 5e and f). The most common arrangement consists of hydrogen-bonded parallel chevrons of hexanol molecules (Fig. 5e), a result previously observed both experimentally and computationally in vacuum[9,48–50]. Within this molecular layer, the oxygen groups are separated by an average distance of 1.52 ± 0.01 nm, coinciding with the features observed experimentally in Fig. 5a. This remarkable agreement validates both the experimental and computational results, bearing in mind the differences in hexanol concentration between experiments and theory. The simulations represent an extreme case where the hexanol concentration is far larger than the 0.7% experimental bulk concentration. This is needed to compensate for the limited time and size of the simulation box. We therefore don't expect the experimental observations to match the simulations over the whole interface due to other possible arrangements and kinetic traps at lower hexanol and methanol concentration.

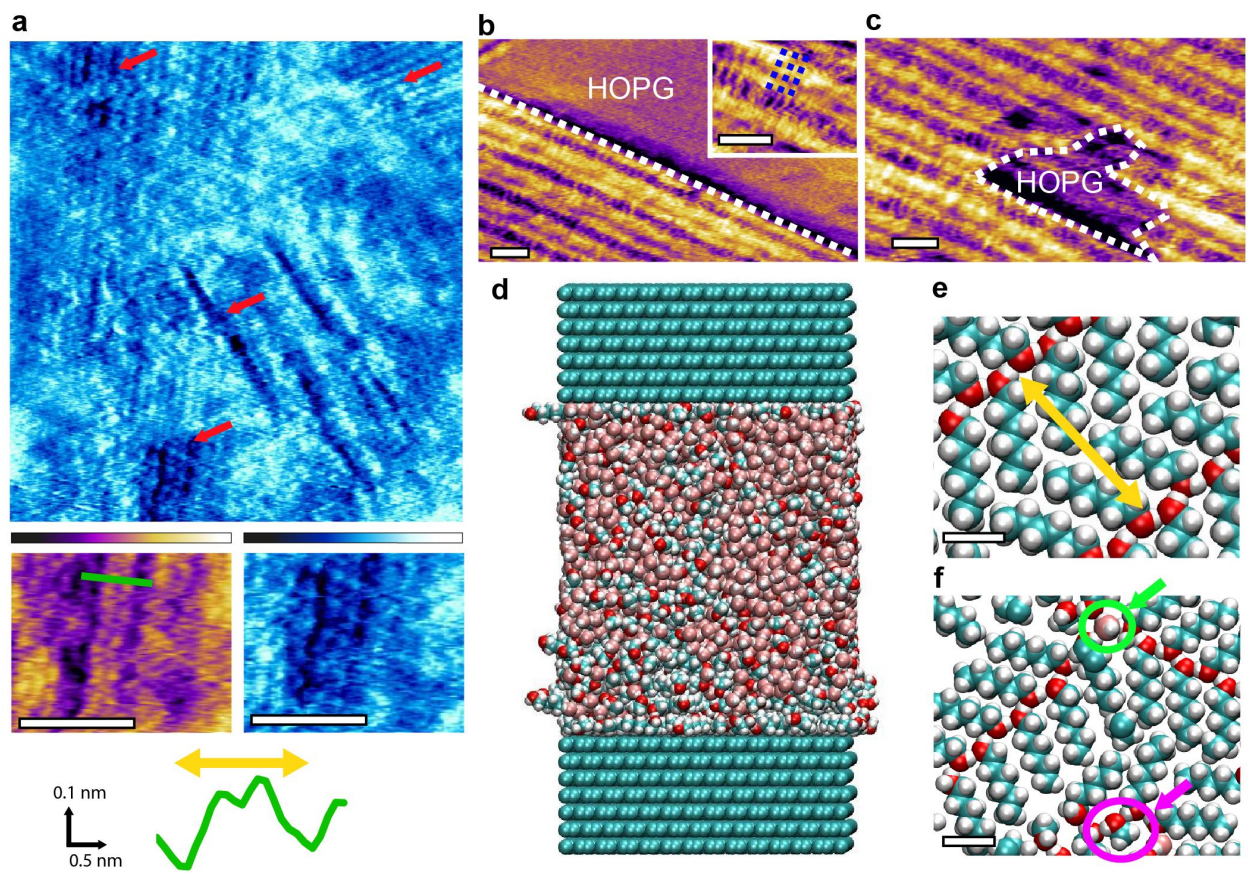

**Figure 5**: Molecular assemblies forming at the surface of HOPG in water-hexanol mixtures. High-resolution AFM images in a water solution containing 0.7% hexanol (saturation) reveal multiple different features (red arrows). Some of the features (inset below (a)) exhibit a periodicity of 1.55 ± 0.05 nm (yellow arrow on green profile). Other typical structures include rows with clean edges (b) (outlined by dashed white line) that exhibit a perpendicular substructure with periodicity of 0.89± 0.08 nm (blue dashed lines in inset). When repeatedly scanning the same area, the AFM creates gaps within the monolayer (c) (white dashed line), exposing the HOPG surface below. MD-simulations of the system are carried out using a box of 10.07 nm × 5.53 nm × 11.3 nm with periodic boundary conditions (d). The solution comprises ~8,000 molecules composing a 10:45:45 hexanol:methanol:water mixture. Snapshots of the interfacial molecular arrangement (e, f) taken within 0.6 nm of the HOPG surface reveal a hexanol self-assembled monolayer with a periodicity of 1.52 ± 0.01 nm ((e), yellow arrow), comparable to the experimental observation in (a). Water molecules (green circle in (f)) and methanol (pink circle in (f)) are also present in the assembly. The scale bars are 5 nm in (a), (b) and (c), and 0.5 nm in (e) and (f). The purple colour scale bar represents a height variation of 0.8 nm in (a) and 1 nm in (b) and (c). The blue scale bar represents a phase bar represents a phase variation of 15º in (a).

Indeed, simulations show that water and methanol molecules (green and pink circles Fig. 5e and f) are involved in the hydrogen bonded networks. They can remain hydrogen-bonded to hexanol molecules within the structured layer for up to 10 ns with some methanol molecules remaining indefinitely embedded in the network over the timescale of the simulation. The involvement of both methanol and water supports the idea of the interfacial molecular assemblies being stabilised by an extended hydrogen bonded network.

The simulations also reveal an important point: direct molecular-substrate interactions can modulate the formation timescale of supramolecular structures. In a previous MD study of methanol-water mixtures at the interface with HOPG[21], we were not able to access the long timescales associated with group nucleation events. In contrast, the stronger interaction between hexanol and HOPG considerably increases its residence time at the interface rendering the nucleation of ordered structures computationally accessible using the present direct MD simulations.

Overall, the flexibility of interfacial self-assembly through group effects hinges on the weak interactions between individual molecules and the surface of the solid so as to prioritise group interactions between assembling molecules to determine the resulting supramolecular structure. In this study we used HOPG as a solid due to its low cost, chemical stability, atomic flatness, and well-defined nanoscale periodicity which enables the incorporation of guest molecules in supramolecular structures[51]. Additionally, self-assembly on HOPG is particularly relevant to graphene-based nanotechnology, with examples in energy storage[52], photonics[53], and water filtration and ion sieving[54]. However, the results obtained with HOPG can, principle, can be extended to any interface with hydrophobic regions flat enough to enable group self-assembly. To test this hypothesis, we replaced the HOPG with either $MoS_2$ or GrO and exposed the substrates to water-methanol mixtures. The results, presented in Fig. 6 confirm the formation of stable structures comparable to the basic monolayer, albeit with some subtle differences that can be ascribed to the substrates.

MoS$_2$ is mildly hydrophobic[55,56] like HOPG and also exhibits a hexagonal symmetry[55]. When a freshly cleaved MoS$_2$ surface is immersed into a 50:50 mixture of methanol and water highly ordered domains composed of row-like structures immediately appear (Fig. 6a). The domains are orientated at 120° with respect to each other indicating an epitaxial influence, as seen on the HOPG and the row pattern looks very similar to that observed on HOPG. However, using MoS$_2$ instead of HOPG still influences the supramolecular structures and their kinetics. First, the molecular domains prefer an elongated growth with single row progressing longitudinally (see Supplementary Fig. 11 for further information). The structures are particularly stable under the applied force of the scanning tip, even for individual rows. Once nucleated the structures grow relatively slowly compared to on HOPG (Fig. 6b-c) and no structures are visible in pure water even after several hours (Supplementary Fig. 12). This suggests that, unlike HOPG, MoS$_2$ is unable to catalyse methanol from water in ambient conditions[20].

Row-like structures can also be observed on GrO in water-methanol mixtures (Fig. 6d-f), but the roughness of the GrO surface and its chemical inhomogeneity at the nanoscale[57] preclude the formation of highly regular structures. The rows on GrO exhibit some variability in width and periodicity and are not visible everywhere on the surface. Here, single GrO flakes have been deposited onto the HOPG substrate so as to offer a clear comparison with the basic monolayer visible on HOPG (Fig 6A, yellow arrow). The exposed GrO surface can be modelled as identical to that of HOPG, but with additional hydrophilic epoxy, hydroxyl and carboxyl groups randomly distributed across the surface[57]. This renders GrO hydrophilic at the macroscale, but it does not exclude nanoscale hydrophobic graphene domains able to template the monolayer self-assembly. Indeed, selective intercalation of GrO sheets has been reported in water-methanol mixtures, consistent with the presence of a specific molecular arrangement of the liquid[58]. The fact that stable structures are able to form (Fig. 6) suggests that the hydrophilic group can either act as influencers or are simply localised enough for the molecular assembly to bridge between hydrophobic nanodomains[59,60].

## Discussion

The data presented in this work investigates the self-assembly of small molecules at interfaces through group effect, without relying on specific or covalent bonds. Group-based self-assembly of small molecules can be achieved using simple systems (here water and methanol) where the interactions between molecules in the bulk solution is strong enough to enable their self-assembly into supramolecular structures once stabilised at an interface. The self-assembly can be significantly influenced through external stimuli, with three main routes available here: (1) the addition of small quantities of influencer molecules such as salts and other hydrogen bonding molecules to modulate both the morphology of the interfacial assemblies and their interactions with the local environment; (2) varying the ratio of alcohol to the other components within the solution; and (3) altering the alcohol-substrate interactions also provide control on the supramolecular assemblies. These three routes can also function in conjunction with one another. For example, ternary 1-propanol-methanol-water mixtures induce the nucleation of multiple well-ordered features characteristic of each component (Supplementary Fig. 9). These strategies make it is possible to create well-reproducible and, to an extent, predictable supramolecular assemblies. The key is to vary the parameters progressively, here using alcohols similar to methanol in molecular structure, symmetry and chemical properties, so as to identify the main evolving trends. Structures with linear features and well-defined but varying periodicities could be consistently created, starting from the basic methanol-water system. The adjunction of influencers tends to induce more dramatic changes which can often be rationalised considering the molecular structure of the influencer. For example, the right-angle symmetry of the sophisticated assembly obtained in presence of phosphate ions is likely due to the tetragonal structure of the ion. The results on MoS$_2$ and GrO indicate a significant degree of flexibility of the group-based strategy which could be extended to a wide range of surfaces, including macroscopically hydrophilic surfaces provided their surface exhibits sufficient nanoscale hydrophobic domains. Further work is however needed to fully explore this idea; the results on GrO are less clear than on the other substrates but control experiments conducted in pure water do not show comparable row-like features (Supplementary Fig. 10). The possibility of extending the concept of group-based self-assembly to other hydrogen bonding small molecules will also require systematic investigation, starting with molecular systems such as ketones, amides and aldehydes.

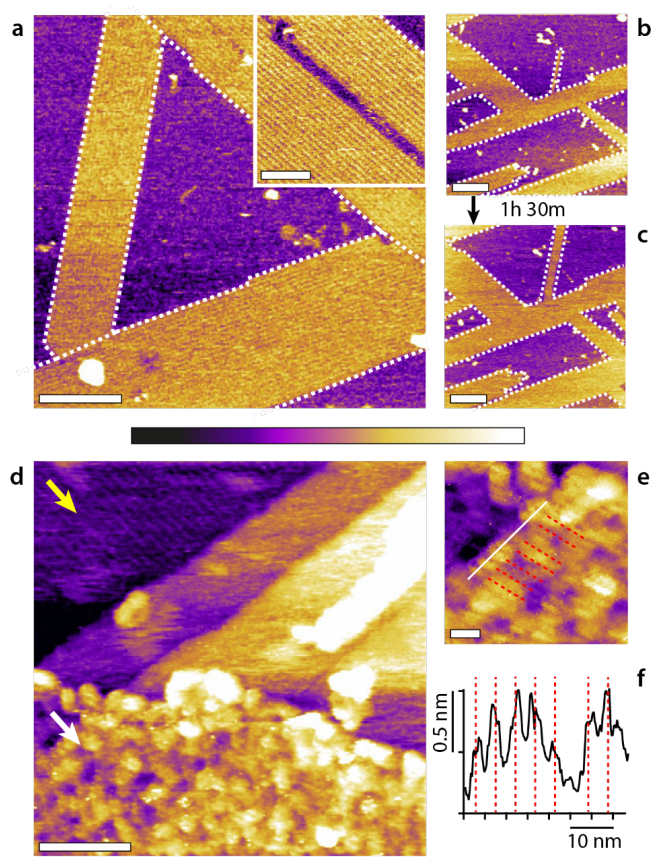

**Figure 6:** Molecular assemblies forming on MoS$_2$ and GrO in a 50:50 methanol-water mixture. On MoS2 (a-c), self-assembled domains with row patterns develop epitaxially (a, white dashed line outline) orientated at 120° with respect to each other. The rows are clearly defined (inset) with a spacing of 5.45 ± 0.05 nm. The row structures grow slowly, by 24.6 ± 0.1 % over 1 h 30 min (b to c), through a 'fingering' mechanism were existing rows tend to elongate longitudinally. The surface would have been fully covered over the same time period with HOPG[21]. The self-assembly on GrO (d-f) is less obvious due to the intrinsic roughness of the GrO surface (d, white arrow) compared to the underlying HOPG substrate where rows are clearly visible (d, yellow arrow). Longitudinal features are however visible at high magnification (e, equidistant red dashed lines corresponding to average separation of 4.5 ± 0.5 nm) with a section (white line) showing clear periodic maxima (d) (see Supplementary Fig. 12 for controls). The scale bars are 100 nm (a), 50 nm (a, inset), 200 nm (b, c), 50 nm (a), and 10 nm (e). The purple scale bar represents a height variation of 1 nm in (a-c), 0.5 nm (a, inset), 2 nm (d) and 0.8 nm (e).

## Conclusions

In this work we explore the concept of group-based self-assembly of small molecules at solid-liquid interfaces. The main difference with standard self-assembly is the fact that the molecules do not significantly interact with the solid and would not durably reside at the interface when isolated. Instead, strong intermolecular interactions allow the molecules to work in group, nucleating ordered structures large enough to remain attached to the solid which then stabilised the system. The fact that individual molecules interact weakly with the solid has one key consequence: the resulting supramolecular assembly can be dramatically influenced by small amounts of foreign molecules or simply by changing the molecular ratios between the main assembling molecules to achieve multiple distinct structures. The idea is illustrated here using water-alcohol mixture spiked with common small molecules to create a wide range of stable supramolecular assemblies at the interface with HOPG at room temperature. These structures can in turn modify the solvation properties of the solid on which they assemble.

Given the diminished importance of specific surface-liquid interactions, group-based self-assembly can in principle occur on many hydrophobic surfaces, here exemplified with MoS$_2$ and GrO that exhibits nanoscale hydrophobic domains. Additionally, the concept may be applied to many other small molecule systems where the molecules are able to form hydrogen bonds and weakly interact with a surface. Further investigations are needed to fully explore the concept's applicability and limitations across different systems and derive a deeper understanding of the molecular details of the structures created, but the concept's simplicity and the high degree of flexibility opens new research avenues for nanotechnology.

## Materials and Methods

**Sample preparation.** All the solutions were prepared with ultrapure water (AnalaR NORMAPUR ISO 3696 Grade 3, VWR Chemicals, Leicestershire, UK). The alcohols used were: HPLC-grade methanol with a purity of ≥99%, HPLC absolute ethanol without additive A15 o[1] with a purity of ≥99.8%, 1-propanol anhydrous with a purity of ≥99.7%, 2-propanol anhydrous with a purity of ≥99.5% and 1-hexanol anhydrous with a purity of ≥99% (all from Sigma-Aldrich, Dorset, UK). The potassium chloride, disodium phosphate and PBS used all had a purity of ≥99% (all from Sigma-Aldrich, Dorset, UK). In a typical experiment, a liquid droplet (~200 µL) of solution was deposited on a freshly cleaved HOPG or MoS$_2$ substrate (both from SPI supplies, West Chester, PA, USA) mounted on a stainless-steel disk. In all cases the substrates were baked at >120 ºC for 20 minutes to remove any contaminants[61] before immediately depositing the droplet. The GrO was synthesised from graphite powder using a modified Hummers method[62], presented in detail elsewhere[63]. To settle the flakes on the HOPG surface a droplet of 1 g/L GrO was deposited on the HOPG and left for 5 minutes before being rinsed with ultrapure water. After the rinsing the methanol water droplet was added in a similar manner to the other experiments on HOPG and MoS$_2$.

**Amplitude Modulation Atomic Force Microscopy.** High-resolution imaging was conducted in amplitude modulation mode in a sealed environment using a commercial Cypher ES AFM (Asylum Research, Santa Barbara, USA) equipped with temperature control and photothermal drive. The sealing of the AFM cell limits the evaporation of the alcohol. The cantilevers (Arrow UHF-AUD, Nanoworld, Neuchatel, Switzerland) had a spring constant of ~1.95 nN/nm (from thermal spectrum calibration) and a resonance frequency of ~430 kHz in liquid. They were cleaned by immersion in ultrapure water before imaging. All parts of the AFM in direct and indirect contact with the solution (cantilever holder, imaging chamber) where thoroughly cleaned with ultrapure water prior to imaging. After washing, the stage was heated to 120 ºC for 20 minutes to evaporate possible substances from previous experiments. All the samples were images at 20.0 ± 0.1 °C except for the results presented in Figs 4(b-c) which were acquired at 30 and 35 °C respectively in an attempt to encourage novel molecular assemblies[21].

**Frequency Modulation Atomic Force Microscopy.** The measurements taken in FM-AFM were acquired using a home-built system[35] with an ultra-low noise cantilever deflection system[64,65]. The AFM head is controlled by a commercially available AFM controller (ARC2, Asylum Research). The tips used in Fig. 2(a-c) and Fig. 3 were AC-55 (Olympus, Tokyo, Japan) with 15 nm of silicon coating (K575XD, Emitech) to improve the stability and reproducibility of the images[66]. The tip quality factor, resonance frequency and spring constant were approximately $Q \approx 12$, $f_o \approx 1.2$ MHz and $k \approx 85$ N/m respectively. A softer cantilever, 15 nm Si coated NCH-AUD (Nanoworld), was needed to obtain stable 3D images in the methanol water system (Fig. 2d and e) where $Q \approx 7$, $f_o \approx 150$ kHz and $k \approx 13.5$ N/m. No temperature control was possible using this system so all samples were at room temperature.

**Molecular Dynamics Simulations.** The simulations were performed using the molecular dynamics package GROMACS

version 2016 [67]. The alcohols molecules and HOPG system were described by the all atom OPLS force field[68]. The water was described by the TIP4P model[69]. The system was a NVT ensemble maintained at 298 K using a velocity rescale thermostat [70] with a coupling time of 0.5 ps. During the simulations, the HOPG atoms were not allowed to move. All simulations were performed with a 0.002 ps timestep. Prior to use, the liquid box was equilibrated for 5 ns in an NPT ensemble at 1 bar and 298 K using the same thermostat and a Parrinello-Rahman barostat[71,72] with a coupling time of 1 ps. After combining with the HOPG, the system was equilibrated for a further 5 ns before the main simulation was performed for a further 16 ns.

## Conflicts of interest

There are no conflicts to declare.

## Acknowledgements

**General:** We are grateful to Dr Jing Zhong (Harbin Institute of Technology) for kindly providing the GrO.
**Funding:** This work was funded by the EPSRC through the Soft Matter and Functional Interfaces CDT (SOFI-CDT), Grant Reference No. EP/L015536/1; World Premier International Research Center Initiative (WPI), MEXT, Japan; JSPS KAKENHI Grant Number 16H02111; and JST Mirai-Project (No. 18077272).